\documentclass[twocolumn,nofootinbib,amsmath,amssymb,a4paper]{revtex4}

\usepackage{graphicx}
\usepackage{dcolumn}
\usepackage{bm}

\def\GV{G_{\mbox{\tiny V}}}
\def\F{{\cal F}}
\def\DRV{\Delta_{\mbox{\tiny R}}^{\mbox{\tiny V}}}
\def\GF{G_{\mbox{\tiny F}}}

\begin{document}

\title{The Status of $V_{ud}$}

\author{J.C. Hardy}
 \email{hardy@comp.tamu.edu}
\affiliation{Cyclotron Institute, Texas A\&M University, College Station, TX 77843}

\begin{abstract}

The best value for $V_{ud}$ comes from superallowed $0^+\rightarrow0^+$ nuclear $\beta$
transitions, of which thirteen have now been measured with high precision.  The current
status of these measurements is described, and the result compared with that of measurements
from neutron decay and pion $\beta$ decay.  Future prospects for improvement are discussed.

\end{abstract}

\maketitle

\section{Introduction}

The up-down quark mixing element, $V_{ud}$, of the Cabibbo Kobayashi Maskawa (CKM) matrix
can be determined via three $\beta$-decay routes: superallowed $0^{+}\!\rightarrow\!0^{+}$ nuclear
decays, neutron decay and pion decay.  Historically, the nuclear decays have yielded the most
precise measurements, but the resultant value for $V_{ud}$ was often thought to be severely
limited in accuracy by the calculated nuclear-structure-dependent corrections that were
required to extract it from the data.  The neutron measurements, though free of nuclear
corrections, are experimentally challenging and have therefore been less precise, not to
mention occasionally inconsistent with one another.  Pion $\beta$ decay, being a $10^{-8}$
decay branch, is even more challenging experimentally and has produced larger
uncertainties for $V_{ud}$ than either of the other two approaches.

All three decay modes require that small radiative corrections be applied to the primary
experimental data in the process of obtaining a value for $V_{ud}$, and naturally there are
uncertainties associated with these calculated corrections.  In fact, in the case of the
superallowed nuclear decays, the experimental uncertainties have become so well controled
that it is these theoretical uncertainties that dominate the uncertainty quoted on $V_{ud}$.
However, it may be a surprise to some readers that the nuclear-structure-dependent correction
is {\it not} the main contributor.  Instead, the dominant theoretical uncertainty originates
from the so-called inner radiative correction, which is a correction that is common
to all three decay modes and, unless it is improved, will ultimately limit the precision
with which $V_{ud}$ can be determined by any route.

As more and more superallowed $0^{+}\!\rightarrow\!0^{+}$ nuclear transitions are measured with high
precision, the nuclear-structure-dependent corrections continue to prove their validity.  The
calculations themselves are based on well-established nuclear structure information derived
from nuclear measurements that are totally independent of the superallowed decay experiments.  The
magnitude of the calculated correction for each transition, though always less than 1.5\%, differs
considerably from transition to transition.  The measured superallowed transition strengths, which
are an order of magnitude more precise than that, actually reproduce these predicted differences and, 
as a result, lead to completely consistent values of $V_{ud}$.  All evidence points to the
nuclear-structure-dependent corrections being completely reliable within their quoted uncertainties.

\section{Superallowed nuclear beta decay}

Beta decay between nuclear analog states of spin-parity, $J^{\pi}=0^+$, and isospin, $T=1$, is a
pure vector transition and is nearly independent of the nuclear structure of the parent and daughter
states.  The measured strength of such a transition -- expressed as an ``$ft$ value" -- can then
be related directly to the vector coupling constant, $\GV$ with the intervention of only a few
small ($\sim$1\%) calculated terms to account for radiative and nuclear-structure-dependent
effects. Once $\GV$ has been determined in this way, it is only another short step to
obtain a value for $V_{ud}$, the up-down mixing element of the Cabibbo-Kobayashi-Maskawa
(CKM) matrix. 

\begin{figure*}[t]
  \includegraphics[width=14.7 cm]{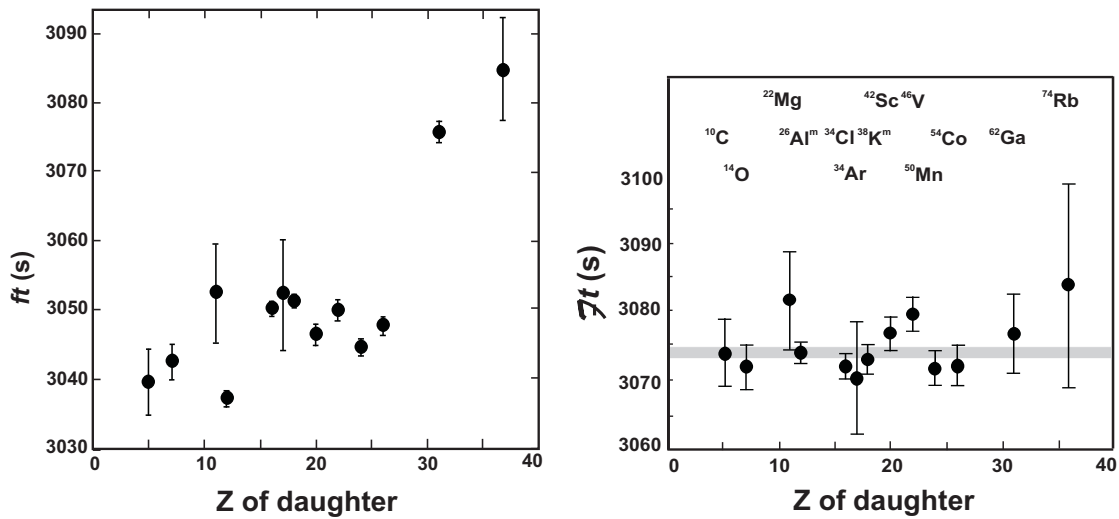}
  \caption{\label{fig1} Results from the 2005 survey \cite{Ha05} updated with more recent published results \cite{Sa05,
To05, Er06a, Er06b, Bo06, Ba06, Ia06, Hy06}.  The uncorrected $ft$ values for the thirteen best known superallowed
decays (left) are compared with the same results after corrections have been applied (right).  The grey band in the
right-hand panel is the average $\F t$ value, including its uncertainty.}
\end{figure*}

In dealing with these decays, which are referred to as superallowed, it is convenient to combine
some of the small correction terms with the measured $ft$-value and define a ``corrected"
$\F t$-value.  Thus, we write \cite{Ha05}
\begin{equation}
\F t \equiv ft (1 + \delta_R^{\prime}) (1 + \delta_{NS} - \delta_C ) = \frac{K}{2 \GV^2 
(1 + \DRV )}~,
\label{Ftconst}
\end{equation}
where $K/(\hbar c)^6 = 2 \pi^3 \hbar \ln 2 / (m_e c^2)^5 = 8120.271(12) \times 10^{-10}$ GeV$^{-4}$s; 
$\delta_C$ is the isospin-symmetry-breaking correction and $\DRV$ is the transition-independent part
of the radiative correction -- the ``inner" radiative correction.  The terms $\delta_R^{\prime}$ and
$\delta_{NS}$ comprise the transition-dependent part of the radiative correction, the former being a
function only of the electron's energy and the $Z$ of the daughter nucleus, while the latter, like
$\delta_C$, depends in its evaluation on the details of nuclear structure.  From this equation, it
can be seen that a measurement of any one of these superallowed transitions establishes an individual
value for $\GV$; moreover, if the Conserved Vector Current (CVC) assertion is correct that $\GV$ is
not renormalized in the nuclear medium, all such values -- and all the $\F t$-values themselves -- 
should be identical within uncertainties, regardless of the specific nuclei involved.

This assertion of CVC can be tested and a value for $\GV$ obtained with a precision considerably
better than 0.1\% if experiment can meet the challenge, since the four small corrections terms only
contribute to the overall uncertainty at the 0.03\% level.  As it turns out, experiment has exceeded
that goal, leaving theory currently as the dominant contributor to the uncertainty.

The $ft$-value that characterizes any $\beta$-transition depends on three measured quantities: the
total transition energy, $Q_{EC}$; the half-life, $t_{1/2}$, of the parent state; and the branching ratio,
$R$, for the particular transition of interest.  The $Q_{EC}$-value is required to determine the statistical
rate function, $f$, while the half-life and branching ratio combine to yield the partial half-life, $t$.
In 2005 a new survey of world data on superallowed $0^{+}\!\rightarrow\!0^{+}$ beta decays was
published \cite{Ha05}.  All previously published measurements were included, even those that were based on
outdated calibrations if enough information was provided that they could be corrected to modern standards.  In
all, more than 125 independent measurements of comparable precision, spanning four decades, made the cut.  
Another 8 publications, with data that can also be incorporated, have appeared \cite{Sa05, To05, Er06a, 
Er06b, Bo06, Ba06, Ia06, Hy06} in the two years since the survey was closed and we have now obtained $ft$
and corrected $\F t$ values for the combined data\footnote{In calculating the $\F t$ values, we have used
slightly different values for $\delta_R^{\prime}$ and $\delta_{NS}$ than those used by us previously \cite{To02}.
This is because the recent improvements in $\DRV$ by Marciano and Sirlin \cite{Ma06} isolated some
transition-dependent components which were more appropriately included in these other terms.  Although the
average value of $\F t$ is thus changed by $\sim$0.3\%, a corresponding change in $\DRV$ leaves the central
value of $V_{ud}$ essentially unchanged.}.  These updated results for the thirteen most precisely
known transitions are shown in Fig.~\ref{fig1}.

By comparing the left and right panels in Fig.~\ref{fig1}, we can see that the calculated corrections act
very well to remove the considerable scatter that is evident in the former but missing in the latter.  Since
$\delta_R^{\prime}$ has essentially the same value for all transitions other than those from $^{10}$C and
$^{14}$O, the removal of the scatter must be attributed to the calculated nuclear-structure-dependent
corrections, $\delta_{NS}$ and $\delta_C$, that have been applied in the derivation of $\F t$.  The
calculation of these correction terms \cite{To02} was based on nuclear-structure models that were solidly
grounded in a wide variety of independent nuclear measurements, none involving superallowed $\beta$ decay. 
Thus the consistency of the $\F t$ values is a powerful validation of those calculated corrections.  All
thirteen $\F t$ values form a statistically consistent set -- with a normalized chi-square of 0.9 -- over a
wide range of nuclear masses, from $A$=10 to $A$=74, a conclusion that is entirely consistent with CVC
expectation.  The average of all thirteen cases is $\overline{\F} t$ = 3073.9(8)s.  Since $\F t$ is
inversely proportional to the {\it square} of $\GV$, this result confirms the constancy of the latter to
1.3 parts in $10^4$, the tightest limit ever set.

The 2005 survey results were also used to set a limit on any possible contribution from scalar currents.  The
presence of a scalar current -- induced or fundamental -- would manifest itself as a curvature, either upwards
or downwards, in the $\F t$-value line at low Z.  There is no hint of any such curvature in the right panel
of Fig.\ref{fig1}, and a careful analysis of the survey results set a limit \cite{Ha05} on the scalar relative
to the vector current of $|C_S/C_V| \leq 0.0013$, again the tightest limit ever set.

With a mutually consistent set of $\F t$ values, no scalar currents, and the test of CVC passed, one can
confidently proceed to determining the value of $\GV$ and, from it, the up-down element of the CKM matrix
via the relation $V_{ud} = \GV/\GF$, where $\GF$ is the well known \cite{PDG06} weak-interaction constant for
purely leptonic muon decay.  From the $\F t$-value data in Fig.~\ref{fig1} and the recently improved calculation
of $\DRV$ \cite{Ma06} we obtain the result,
\begin{equation}
V_{ud} = 0.97378(27).~~~~~~~~~~~~~~{\rm [nuclear]}
\label{nuclear}
\end{equation}
This number can be compared with our
previous value, quoted in the 2005 survey, $V_{ud}$ = 0.97380(40).  The new result is completely consistent
with the earlier one, but it has a considerably reduced uncertainty -- mostly due to the improvement in the
calculated ``inner" radiative correction, $\DRV$.

\begin{figure}[t]
  \includegraphics[width=0.47\textwidth]{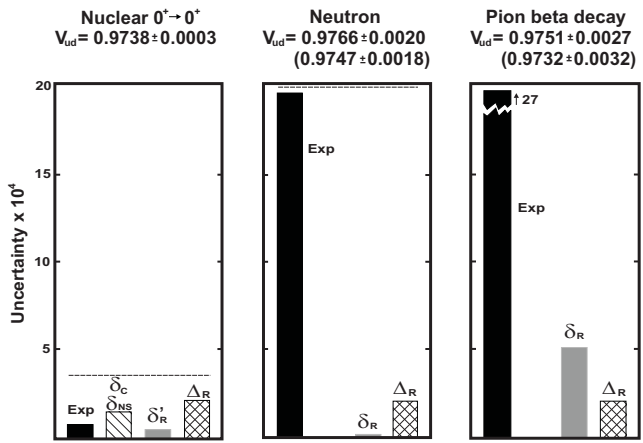}
  \caption{\label{fig2} Error budgets for the three different methods to determine $V_{ud}$, illustrating
the relative importance of experimental uncertainties and theoretical ones.  The unbracketed values of
$V_{ud}$ correspond to the the results shown in numbered equations in the text; bracketed values
are alternative results obtained under different conditions also described in the text.}
\end{figure}

The ``error budget" leading to the uncertainty quoted for $V_{ud}$ in Eq.\,\ref{nuclear} is illustrated in
the left panel of Figure \ref{fig2}.  The largest contribution to the uncertainty comes from $\DRV$
($\pm$0.00018), while the nuclear-structure-dependent corrections contribute $\pm$0.00015 and experimental
uncertainties only $\pm$0.00007.

\section{Neutron and pion beta decays}

On the one hand, free neutron decay has an advantage over nuclear decays since there are no nuclear-structure-
dependent corrections to be calculated.  On the other hand, it has the disadvantage that it is not purely
vector-like but has a mix of vector and axial-vector contributions.  Thus, in addition to a lifetime measurement,
a correlation experiment is also required to separate the vector and axial-vector pieces.  Both types of
experiment present serious experimental challenges.  The value of $V_{ud}$ is determined from the expression

\begin{equation}
V_{ud}^2 = \frac{K / \ln 2}{\GF^2 (1 + \DRV )
(1 + 3
\lambda^2 )
f (1 + \delta_R ) \tau_n } ,
\label{Vud2n}
\end{equation}

\noindent where $\lambda$ is the ratio of axial-vector and vector effective coupling constants, $\tau_n$ is
the mean life for neutron decay, $f$ is the statistical rate function and $\delta_R$ is the
transition-dependent radiative correction evaluated for the case of a neutron.  Both $\lambda$ and $\tau_n$
must be determined from experiment.

In 2003, a survey of world data for neutron $\beta$ decay was published \cite{To03}, in which the same
citeria were used as for the superallowed nuclear beta-decay measurements:~{\it i.e.}~all measurements were
retained unless they had been withdrawn by the original authors or could be rejected on objective grounds.
Since that time, there have been two new mean-life measurements \cite{Ni05,Se05} that fit within the
criteria for inclusion.  Strikingly, one of these measurements, by Serebrov {\it et al.}~\cite{Se05}, disagrees
with the previous world average by more than six standard deviations.  However, without any objective grounds for
rejecting this discrepant result -- or, alternatively, rejecting all preceeding measurements -- we have evaluated
the world average with all data included.  The result for the neutron mean life is then $\tau_n$ = 882.0(15) s
(with $\chi^2$/N = 5.4!); and for the coupling-constant ratio is $\lambda$ = -1.2690(28) (with $\chi^2$/N = 2.6).

Inserting these average experimental results for $\tau_n$ and $\lambda$ into Eq. \ref{Vud2n}, we determine
$V_{ud}$ to be 
\begin{equation}
V_{ud} = 0.9766(20),~~~~~~~~~~~~~~{\rm [neutron]}
\label{neutron}
\end{equation}
a value with much larger uncertainty than the nuclear result and just outside of one standard deviation away
from it.  It is also worth noting, though, that in considering the neutron mean life, the Particle Data
Group has taken a different approach \cite{PDG06} and has completely rejected the Serebrov {\it et al.}~\cite{Se05}
result simply because it does disagree significantly with previous measurements.  If we do the same thing, 
then we obtain $V_{ud}$ = 0.9747(18), which fully overlaps the nuclear result but still has an uncertainty that
is nearly 7 times larger.

The error budget for the result in Eq. \ref{neutron} appears in the center panel of Figure \ref{fig2}.  Unlike
the nuclear result in the left panel, this uncertainty is completely dominated by experiment.  Nevertheless, it
is important to recognize that the contribution from $\DRV$ is the same for the neutron as it is for the nuclear
decays.  Even as the neutron's experimental results are improved, the value of $V_{ud}$ obtained from them will
ultimately be limited in precision at very nearly the same level that the nuclear result has currently reached.

Like neutron decay, pion beta decay has an advantage over nuclear decays in that there are no
nuclear-structure-dependent corrections to be made.  It also has the same advantage as the nuclear decays in
being a purely vector transition, in its case $0^{-} \rightarrow 0^{-}$, so no separation of vector and
axial-vector components is required.  Its major disadvantage, however, is that the beta-decay mode, 
$\pi^{+} \rightarrow \pi^0 e^{+} \nu_e$, is a $\sim 10^{-8}$ branch of the total pion decay.  This results
in severe experimental limitations.  For the pion beta decay, the value of $V_{ud}$ is determined from
the expression
\begin{equation}
V_{ud}^2 = \frac{K /\ln 2}{2 \GF^2 (1 + \DRV )
f (1 + \delta_R) \tau_{\pi} / BR },
\label{Vud2pi}
\end{equation}
where $\tau_{\pi}$ and $BR$ are the mean life and branching ratio for pion beta decay, $f$ is the statistical
rate function and $\delta_R$ is the transition-dependent radiative correction evaluated for the case of the
pion.  The current world average \cite{PDG06} for the mean life is $\tau_{\pi}$ = 2.6033(5) $\times 10^{-8}$ s.
As for the beta-decay branching ratio, the world average is now dominated by the recent PIBETA measurement
\cite{Po04}, which determined it relative to the much stronger $\pi^{+}\!\rightarrow\! e^{+} \nu_e(\gamma)$
branch and improved its experimental precision by more than a factor of 6.  The latter branch has been
determined experimentally but can also be determined -- more precisely, in fact -- by theoretical calculation
\cite{PDG06}.

With the theoretical value $BR(\pi^{+}\!\rightarrow e^{+}\!\nu_e(\gamma))$ = 1.2352(5) $\times 10^{-4}$, the
PIBETA measurement of the pion beta-decay branching ratio becomes 1.040(6) $\times 10^{-8}$.  Equation
\ref{Vud2pi} then yields the result
\begin{equation}
V_{ud} = 0.9751(27),~~~~~~~~~~~~~~{\rm [pion]}
\label{pion}
\end{equation}
which agrees well with the nuclear result but with 10 times the uncertainty.  If, instead, the experimental
measurement had been used for the $\pi^{+} \rightarrow e^{+} \nu_e(\gamma)$ branching ratio, then $V_{ud}$
= 0.9732(32), which also is consistent with the nuclear result but with an even larger uncertainty.

\section{Conclusions}

As of now, superallowed $0^+\rightarrow0^+$ nuclear $\beta$ decay clearly dominates in the determination of
$V_{ud}$.  A weighted average of the results quoted in Eqs.~\ref{nuclear}, \ref{neutron} and \ref{pion},
yields a result that differs only in the fifth place of decimals from the nuclear result alone; and, considering
the ambiguities present in the other decays, especially in that of the neutron, it seems best for the time being
to rely simply upon the unaveraged nuclear result.  This result is not limited by experimental uncertainties
but by uncertainties originating in the theoretical corrections applied to the data.  Even so, the largest
theoretical uncertainty is not due to the nuclear-structure-dependent corrections but rather to the ``inner"
radiative correction, $\DRV$, which actually is common to all three $\beta$ decays: nuclear, neutron and pion.
Clearly it would be beneficial to have confirmation of the nuclear $V_{ud}$ by results of comparable
precision from the other two $\beta$-decay modes, but one should not anticipate that either of the latter
will actually surpass the nuclear result in the near future.

Although experiment is not currently limiting the precision of the nuclear $V_{ud}$, there is considerable
activity at the moment in improving experimental precision and extending measurements to previously
unstudied superallowed transitions.  The approach is best illustrated by Fig.\,\ref{fig1} and the observation that
the calculated corrections are validated by their success in removing the $ft$-value scatter.  Improvements
in experimental precision would test the calculations even more severely, as would new examples of
$0^{+}\!\rightarrow\!0^{+}$ transitions specifically selected for having {\it larger} calculated corrections.  
The reasoning is that if the $ft$ values measured for cases with large calculated corrections also turn
into corrected $\F t$ values that are consistent with the others, then this must verify the calculations'
reliability for the existing cases, which have smaller corrections.  In fact, the cases of $^{34}$Ar, $^{62}$Ga
and $^{74}$Rb were chosen for this reason and their consistency has already led to considerably increased
confidence in the calculated corrections.

It is reasonable to expect that there will be some further improvement in the nuclear value for $V_{ud}$ as
a result of this campaign to test the nuclear-structure-dependent corrections.  However, really significant
improvements must await a more precise calculation of $\DRV$. 

\begin{acknowledgments}

Much of the work reported here was done in collaboration with I.S. Towner.  It was supported financially by
the U.S. Department of Energy under Grant No.~DE-FG03-93ER40773 and by the Robert A. Welch Foundation under
Grant No.~A-1397.

\end{acknowledgments}

\end{document}